%% file: main.tex
\documentclass[a4paper,11pt]{article}
\usepackage{pos}

\usepackage{orcidlink}
\usepackage{slashed}
\usepackage{mathtools}
\usepackage{cleveref}
\usepackage{comment}
\usepackage{graphicx}
\usepackage{caption}
\usepackage{subcaption}
\usepackage{tabularx}
\usepackage{amsmath}
\usepackage{amsfonts}
\usepackage{listings}

\newcommand{\ii}{\mathrm{i}}
\newcommand{\dd}{\mathrm{d}}
\PassOptionsToPackage{export}{adjustbox}

\def\pmtt{\texttt{+\!\!\!\raisebox{-.48ex}-}}

\def\Rindex{R}

\newcommand{\planar}{\text{P}}
\newcommand{\nonplanar}{\text{NP}}
\newcommand{\floopsza}{N_{Z\gamma}}

\usepackage{multirow}
\usepackage{float}
 \usepackage{makecell}
 \newcommand{\ccline}[1]{%
    \cline{#1}%
    \noalign{\vskip-\doublerulesep}
    \cline{#1}%
    \noalign{\vskip\doublerulesep}
}
\def\twoloopfloopsnonplanar{\widehat{F}^{(2,\floopsza),R}_\nonplanar}

\def\twoloopfloopsplanarone{\widehat{F}^{(2,\floopsza),R}_{\planar_1}}
\def\twoloopfloopsplanartwo{\widehat{F}^{(2,\floopsza),R}_{\planar_2}}
\def\fulltwoloopfloops{F^{(2,\floopsza)}}
\def\twoloopfloopsuv{{R}_{\rm UV} \widehat{F}^{(2,\floopsza)}}

\def\twoloopfloopsnonplanarcs{\widehat{\sigma}^{(2,\floopsza),R}_\nonplanar}
\def\twoloopfloopsplanarcs{\widehat{\sigma}^{(2,\floopsza),R}_\planar}
\def\twoloopfloopsplanaronecs{\widehat{\sigma}^{(2,\floopsza),R}_{\planar_1}}
\def\twoloopfloopsplanartwocs{\widehat{\sigma}^{(2,\floopsza),R}_{\planar_2}}
\def\fulltwoloopfloopscs{\sigma^{(2,\floopsza)}}
\def\twoloopfloopsuvcs{{R}_{\rm UV} \widehat{\sigma}^{(2,\floopsza)}}

\def\ttbf{\ttfamily\fontseries{b}\selectfont}
\def\ttlf{\ttfamily\fontseries{l}\selectfont}

\usepackage{xspace}

\tikzset{
cut/.style={
    opacity=0.4,
    line width=2pt,
    shorten <=-15pt,
    shorten >=-15pt}
}
\usepackage[compat=1.1.0]{tikz-feynman}
\def\ttbf{\ttfamily\fontseries{b}\selectfont}
\def\ttlf{\ttfamily\fontseries{l}\selectfont}

\usepackage{listings}
\DeclareCaptionLabelFormat{boldlisting}{\textbf{Listing~#2}}
\chardef\MyArticleWithColor=\pdfcolorstackinit page direct{0 g}

\title{Heavy-quark box-loop corrections to $q\bar q \to Z\gamma$ at two loops in QCD}
\ShortTitle{Two-loop heavy-quark box contributions to $q\bar q \to Z\gamma$}

\author*[a,b]{Dario Kermanschah}
\author[a]{Matilde Vicini}

\affiliation[a]{Institute for Theoretical Physics, ETH Zurich, Wolfgang-Pauli-Strasse 27, 8093 Z\"urich, Switzerland}

\affiliation[b]{Rudolf Peierls Centre for Theoretical Physics, Oxford University, Clarendon Laboratory, Parks Road, Oxford OX1 3PU, UK}

\emailAdd{d.kermanschah@gmail.com}
\emailAdd{mvicini@phys.ethz.ch}

\abstract{
We numerically compute the two-loop QCD corrections to $Z\gamma$ production at the LHC mediated by light- and heavy-quark box loops.
The calculation employs the pipeline of refs.~\cite{Kermanschah:2024utt,Kermanschah:2025wlo}, which performs Monte Carlo integration over spatial loop momenta after local subtraction of infrared, ultraviolet, and threshold singularities.
We validate our results for partonic squared matrix elements with massless-quark loops against known benchmarks, extend them to include heavy-quark contributions, and compute the double-virtual corrections to $pp\to Z\gamma$ by performing the loop and phase space integrations simultaneously.
This computation demonstrates the flexibility of the approach in handling both massless and massive final-state bosons, as well as additional mass scales in the loop.
}

\FullConference{17th International Symposium on Radiative Corrections: Applications of Quantum Field Theory to Phenomenology (RADCOR2025)\\
5-10 October 2025\\
Puri, India\\}

\begin{document}
\maketitle

\section{Introduction}
The computation of two-loop multi-leg scattering amplitudes with several external legs and multiple mass scales remains a major challenge in the field.
In previous works~\cite{Kermanschah:2024utt,Kermanschah:2025wlo}, we developed a numerical approach resilient to this level of complexity and demonstrated its feasibility by computing fermionic two-loop corrections to di- and triphoton production, involving up to eight independent kinematic and mass scales.
\par
In these proceedings, we apply our techniques to the process $q\bar{q} \to Z\gamma$ mediated by light and heavy quark box loops.
We exploit the fact that these contributions are kinematically equivalent to those where the $Z$ boson is replaced by an off-shell photon.
Therefore, the framework developed in ref.~\cite{Kermanschah:2025wlo} applies directly to the production of a massless–massive boson pair.
We recompute the two-loop squared matrix elements with massless quark loops and validate them against the analytic results of ref.~\cite{Gehrmann:2011ab}.
In addition, we present new results for the contributions with heavy-quark box loops, for which no analytic benchmarks are currently available.
\par
Our implementation can be directly used for cross section calculations of proton-proton scattering at the Large Hadron Collider (LHC) when combined with real-emission contributions.
We illustrate this by computing the double-virtual corrections, convoluting the squared matrix element with parton distribution functions (PDFs), and performing the Monte Carlo integration over loop and phase space simultaneously.
\par
In parallel work, we demonstrate that the same methods also apply to triangular fermion-loop corrections.
These contributions vanish in massless quantum chromodynamics (QCD) but remain non-zero when heavy top and bottom quarks are present in the loop.
\par
Our approach follows ref.~\cite{Kermanschah:2025wlo} and constructs finite two-loop amplitudes in loop momentum space.
It combines local infrared (IR) factorization at two loops~\cite{Anastasiou:2020sdt,Anastasiou:2022eym,Anastasiou:2024xvk,Anastasiou:2025cvy,Anastasiou:2026kpm} with local ultraviolet (UV) renormalization~\cite{Bogoliubov:1957gp,Hepp:1966eg,Zimmermann:1969jj}, before analytically integrating over the energy components of the loop momenta.
This procedure leads to loop-tree duality representation \cite{Catani:2008xa, Aguilera-Verdugo:2019kbz, Aguilera-Verdugo:2020set,JesusAguilera-Verdugo:2020fsn, Ramirez-Uribe:2022sja, Runkel:2019yrs, Capatti:2019ypt}, or equivalent but numerically more stable expressions arising from (partially) time-ordered perturbation theory \cite{Bodwin:1984hc, Collins:1985ue, Sterman:1993hfp, Sterman:1995fz,Sterman:2023xdj,Aguilera-Verdugo:2020set, Aguilera-Verdugo:2020kzc, JesusAguilera-Verdugo:2020fsn,Ramirez-Uribe:2020hes, Sborlini:2021owe, TorresBobadilla:2021ivx, Bobadilla:2021pvr, Benincasa:2021qcb, Kromin:2022txz, Capatti:2020ytd,Capatti:2022mly, Capatti:2023shz}.
The remaining threshold singularities are treated in accordance with the $\ii\epsilon$ causal prescription by subtracting them using the method of refs.~\cite{Kermanschah:2021wbk,Kermanschah:2024utt,Vicini:2024ecf,Kermanschah:2025wlo}.
This approach improves control and numerical efficiency over contour deformation into the complex plane~\cite{Soper:1999xk,Becker:2011vg,Becker:2012aqa,Buchta:2015wna,Kromin:2022txz,Capatti:2019edf}.
The resulting expression is numerically integrated over spatial loop momenta, and if desired, over phase space, using a multi-channel Monte Carlo method with adaptive importance sampling based on the \textsc{Vegas} algorithm \cite{Lepage:2020tgj,Lepage:1977sw,Hahn:2004fe,havana}.

\section{Calculation}

\input{figures/diags}
We consider the process
\begin{align}
    q(p_1)\, \bar q(p_2) \to \gamma(q_1)\, Z(q_2)
\end{align}
at two loops in QCD, focusing on the contribution mediated by box fermion loops in which a quark of flavor $f$ with mass $m_f \ge 0$ circulates.
The contributing Feynman diagrams with a single charge flow are shown in figure~\ref{fig:diagrams}.
The incoming quarks and outgoing photon are massless, $p_1^2 = p_2^2 = q_1^2 = 0$, while the $Z$ boson is massive, $q_2^2 > 0$.
The coupling to quark vector and axial-vector currents is written as
\begin{align}
    \ii e \gamma^\mu (v_f^X - a_f^X \gamma^5), \qquad X \in \{Z, \gamma\},
\end{align}
where $e=\sqrt{4\pi\alpha}$ is the positron charge.
For the $Z$ boson, the couplings are expressed in terms of the weak mixing angle $\theta_W$, the third component of weak isospin $I^3_f$ and the electric charge $Q_f$,
\begin{align}
    v_f^Z = \frac{I^3_f - 2 s^2 Q_f}{2 s c}, \qquad
    a_f^Z = \frac{I^3_f}{2 s c},
\end{align}
where $s=\sin\theta_W$ and $c=\cos\theta_W$.
The photon couples only to the vector current,
\begin{align}
    v_f^\gamma = -Q_f, \qquad
    a_f^\gamma = 0.
\end{align}
\par
Charge conjugation invariance of the subprocess $gg \to Z\gamma$ implies that fermion loops with three vector and one axial coupling vanish.
Hence, we can drop the axial coupling in the loop and compute the amplitude with vector couplings only, effectively replacing the $Z$ boson with an off-shell photon $\gamma^*$ up to the difference in their vector couplings.
Therefore, the integrands constructed for ref.~\cite{Kermanschah:2025wlo} for fermionic corrections to photon production can be reused here.
\par
We denote the contribution to the two-loop scattering amplitude as
\begin{align}
\label{eq:order}
    \left(\frac{\alpha_s}{4 \pi}\right)^2 C_F T_F \floopsza \delta_{ij} M^{(2,\floopsza)},
\end{align}
where we factored out the strong coupling $\alpha_s$, the color factors
\begin{align}
    C_F=\frac{N^2-1}{N}, \qquad T_F=\frac{1}{2},
\end{align}
with $i$ and $j$ the color indices of the incoming quark, and the quark vector couplings of the $Z$ boson and photon summed over all equal-mass flavors:
\begin{align}
\label{eq:coupling}
    \floopsza = \sum_f e^2 v_f^Z v_f^\gamma.
\end{align}
The tree-level amplitude is
\begin{align}
\label{eq:tree}
    \delta_{ij} e^2 v_q^\gamma\left(v_q^Z M^{(0,V)}+a_q^Z M^{(0,A)}\right),
\end{align}
where we separated the vector and axial-vector contributions and factored out the corresponding couplings.
\par

\subsection{IR and UV subtraction}
The box-fermion loop corrections $M^{(2,\floopsza)}$ are finite in four spacetime dimensions, and neither require renormalization nor depend on the factorization scheme.
However, the individual Feynman diagrams are IR and UV divergent. 
These divergences do not cancel locally in their sum but only globally after loop integration in dimensional regularization.
To enable direct numerical integration in momentum space, we must first render the integrand power-counting finite by introducing local IR and UV counterterms.
\par
We express each Feynman diagram in figure~\ref{fig:diagrams} as
\begin{align}
    G = \int \frac{\dd^d k_1}{\pi^{d/2}}\frac{\dd^d k_2}{\pi^{d/2}} \mathcal{G},
\end{align}
and decompose the integrand in terms of its subgraphs,
\begin{align}
    \mathcal{G} = \mathcal{D}^{\alpha\beta}\mathcal{V}_{\alpha\beta},
\end{align}
where the incoming quark line in Feynman gauge is
\begin{align}
    \mathcal{D}^{\alpha\beta}= i\frac{\overline{v}(p_2)\gamma^\beta(-\slashed{k_1})\gamma^\alpha u(p_1)}{k_1^2 k_{g_1}^2 k_{g_2}^2},
\end{align}
with gluon momenta $k_{g_i}$, and $V_{\alpha\beta}$ denotes the fermion-loop subgraph.
The $i\epsilon$ causal prescription in the Feynman propagators is left implicit.
\par
The integral has divergences associated to longitudinal gluons collinear to the incoming massless quarks.
Suitable IR counterterms were proposed in refs.~\cite{Anastasiou:2020sdt,Anastasiou:2024xvk,Kermanschah:2025wlo}.
Analogous to 
ref.~\cite{Kermanschah:2025wlo}, we subtract the two collinear limits $k||p_i$ using the counterterms
\begin{align}
    \Delta_1 \mathcal{D}^{\alpha\beta}
    &= -i\frac{\overline{v}(p_2)\gamma^\beta u(p_1)}{k_1^2 k_{g_1}^2} \frac{p_1\cdot \zeta}{p_1\cdot p_2} \frac{2k_{g_1}^\alpha}{(k_1-\zeta)^2-\zeta^2}, \\
    \Delta_2 \mathcal{D}^{\alpha\beta}
    &= -i\frac{\overline{v}(p_2)\gamma^\alpha u(p_1)}{k_1^2 k_{g_2}^2} \frac{p_2\cdot \zeta}{p_1\cdot p_2} \frac{-2k_{g_2}^\beta}{(k_1-\zeta)^2-\zeta^2},
\end{align}
where $\zeta$ is an auxiliary vector satisfying $p_1\cdot\zeta\neq0$, $p_2\cdot\zeta\neq0$ and $\zeta^2 \neq 0$, which we choose as $\zeta=p_1+p_2$.
\par
UV divergences from the box-quark loops $V_{\alpha\beta}$ are subtracted with UV counterterms constructed using the $R$-operation, which here corresponds to replacing the fermion propagator
\begin{align}
\frac{i}{\slashed{k}_2+\slashed{p}-m} \quad \mapsto \quad  \frac{i\slashed{k}_2}{k_2^2-M^2},
\end{align}
where $p$ is some momentum shift and $M$ is a UV regulating mass.
We note that the amplitude is independent of $M$ as it is finite.
\par
With these counterterms, each Feynman diagram becomes a finite integral in four dimensions:
\begin{align}
    \widehat{\mathcal{G}}^\Rindex = (1-\Delta_1-\Delta_2)\mathcal{D}^{\alpha\beta} \ (1-\Rindex_\text{UV})\mathcal{V}_{\alpha\beta}.
\end{align}
The Ward identity ensures that the sum of all IR counterterms integrates to zero \cite{Anastasiou:2020sdt,Anastasiou:2024xvk}.
Moreover, the sum of UV counterterms reduces to the finite quantity 
after integrating over $k_2$ in dimensional regularization~\cite{Anastasiou:2020sdt}, leaving a finite one-loop integral over $k_1$, which we also integrate numerically.
\par
Finally, the amplitude can be expressed in terms of individually finite contributions:
\begin{align}
\label{eq:parts}
M^{(2,\floopsza)}=
2\left(
\widehat{G}_{\planar_1}^\Rindex + \widehat{G}_{\planar_2}^\Rindex +
\widehat{G}_{\nonplanar}^\Rindex
\right)
+ \Rindex_\mathrm{UV} M^{(2,\floopsza)},
\end{align}
where the factor of two accounts for the diagrams with reversed charge flow, $\widehat{G}_{\planar_i}$ and $\widehat{G}_{\nonplanar}$ are the finite versions of the planar and non-planar diagrams in figure~\ref{fig:diagrams}, and $\Rindex_\text{UV}M^{(2,\floopsza)}$ denotes the sum of UV counterterms.
\par

\subsection{Loop-energy integration and threshold subtraction}
For each of the finite contributions above in eq.~\eqref{eq:parts}, we symbolically integrate over the energy components of the loop momenta using a \textsc{FORM} \cite{Kuipers:2013pba,Ruijl:2013epa,Davies:2026cci} script.
The script first splits the propagators into linear denominators:
\begin{align}
    \frac{n(q^0)}{q^2-m^2} = \frac{1}{2E}\left(\frac{n(E)}{q^0-E}-\frac{n(-E)}{q^0+E}\right)
\end{align}
where $n(q^0)$ is any linear numerator.
It then applies partial fraction decomposition in the loop energies $k_i^0$, such that only denominators with sums of on-shell energies are introduced, keeping only those terms in which a single simple pole in $k_i^0$ remains, which, if inside the chosen contour, integrates to its residue.
\par
In general, this procedure can generate spurious singularities similar to those in time-ordered perturbation theory, involving sums of more than $n+1$ on-shell energies for an $n$-loop integrand.
Although spurious singularities of this type do not significantly affect numerical stability, we nonetheless checked that they did not arise in the contributions considered here.
The resulting expressions are therefore equivalent to those obtained from partially time-ordered perturbation theory \cite{Sterman:2023xdj}, the cross-free family \cite{Capatti:2022mly}, or causal loop-tree duality representations \cite{Sborlini:2021owe,TorresBobadilla:2021ivx,Bobadilla:2021pvr,Benincasa:2021qcb}.
The practical advantage of our algorithm is that it operates entirely algebraically and does not require knowledge of an underlying graph structure.
\par
The integrand after loop energy integration, which we denote by $\mathcal{I}$, still has threshold singularities, identified by the Cutkosky cuts illustrated in figure~\ref{fig:diagrams}, e.g. the teal cut
\begin{align}
    t: \sqrt{\left(\vec{k}_1-\vec{p}_2\right)^2} + \sqrt{\left(\vec{k}_1+\vec{k}_2+\vec{p}_1\right)^2+m_f^2}+ \sqrt{\left.\vec{k}_2\right.^2+m_f^2} - p_1^0-p_2^0 = 0.
\end{align}
If the quark mass $m_f$ is sufficiently large, i.e. $q_2^2<4m_f^2$ or $s=(p_1+p_2)^2<4m_f^2$ the final-state or $s$-channel cuts through the quark-loops do not admit solutions for real loop momenta.
Each threshold singularity $t$ is then subtracted using a counterterm:
\begin{align}
\label{eq:threshold_ct}
    \mathcal{C}_t = \frac{\text{Res}[\mathcal{I},r=r_t]}{r-r_t}
    \left(\frac{2 r_t}{r+r_t}\right)^{6n-1},
\end{align}
where $r_t$ is the root of $t=0$, and $r=|\mathbf{r}|$ is the magnitude of the $n$ spatial loop momenta $\mathbf{r}=(\vec{k}_1,\dots,\vec{k}_n)+(\vec{s}_1,\dots,\vec{s}_n)$ with origin shifts $\vec{s}_i$.
\par
Here, for the two-loop squared matrix elements, we take $\vec{s}_1 = \vec{p}_2$ in all cases, with $\vec{s}_2 = \vec{0}$ for planar and $\vec{s}_2 = \vec{q}_1/2$ for non-planar contributions.
This choice is generally not suitable across the full phase space, so for the phase-space integrated virtual corrections we keep $\vec{s}_1 = \vec{p}_2$ and adjust $\vec{s}_2$ according to the region.
If $q_2^2 > 4 m_f^2$, we set $\vec{s}_2 = \vec{q}_1/2$ for both contributions.
Otherwise, the planar case uses $\vec{s}_2 = \vec{0}$, while the non-planar employs two channels with sources $\vec{s}_2 = \vec{0}$ and $\vec{s}_2 = \vec{q}_1$, with singular surfaces separated as described in ref.~\cite{Kermanschah:2025wlo}.
All calculations are performed in the rest frame of $p_1 + p_2$.
\par
The residue $\text{Res}[\mathcal{I},r=r_t]$ corresponds to the product of the tree amplitudes separated by the cut.
To avoid introducing a UV divergence, it is multiplied by a suppression function, constructed so that the Cauchy principal value integral vanishes.
In eq.~\eqref{eq:threshold_ct}, this is ensured by antisymmetry under inversion, i.e. 
\begin{align}
f\left(\frac{r_t^2}{r^2}\mathbf{r}\right)=-\left(\frac{r^2}{r_t^2}\right)^{3n}f\left(\mathbf{r}\right),
\end{align}
although alternative UV suppression functions can be used (cf.~\cite{Kermanschah:2021wbk,Kermanschah:2024utt,Kilian:2009wy}).
\par
The dispersive part of the integral is obtained by integrating the subtracted integrand, $\mathcal{I}-\sum_t \mathcal{C}_t$, over the spatial loop momenta, while the absorptive part comes from integrating the sum of residues over the $3n-1$ dimensional space parameterized by $\hat{\mathbf{r}}$, where we can explicitly set the spacetime dimension to $d=4$ and remove the $\ii \epsilon$ causal prescription.
\par
For the process considered here, only the real part of the dispersive integral contributes to the squared matrix element. Therefore, in this case, though not in general, the imaginary part of the absorptive contribution can be neglected.
\par
The integration over the energy components of the loop momenta, as well as the construction of the threshold counterterms is automated in our pipeline, previously used for the calculations in refs.~\cite{Kermanschah:2024utt,Vicini:2024ecf,Kermanschah:2025wlo}.
We use \textsc{Spenso}~\cite{spenso} and \textsc{Symbolica}~\cite{symbolica} to optimize the integrand expressions for fast evaluation.

\section{Numerical results}
\label{sec:results}

\input{figures/table_za_light}

We evaluate the helicity-summed squared matrix element
\begin{align}
\label{eq:matrix-element}
    \fulltwoloopfloops &= 2\,\text{Re} \sum_h M_h^{(2,\floopsza)}\left(M_h^{(0,V)}\right)^*,
\end{align}
where the sum runs over the helicities $h$ of the external particles.
In our framework, it is obtained by multiplying the integrand of the dispersive part of our locally finite amplitude $M^{(2,\tilde{N}_f)}$ with the complex conjugate of the vector-contribution of tree amplitude $M_h^{(0,V)}$, summing over all helicities, and performing the numerical integration over the spatial loop momenta via Monte Carlo sampling.
\par
We neglect the axial-vector coupling in the tree amplitude, as its contribution vanishes in the matrix element due to the absence of four linearly independent vectors to contract with the antisymmetric tensor $\epsilon^{\mu\nu\rho\sigma}$ generated by $\gamma^5$.
Exploiting parity invariance, we sum over only half the helicities and account for the other half with a factor of two.
\par
We compute the squared matrix element for three randomly generated phase-space points, given in listing~\ref{lst:ps_points}, at a center-of-mass energy $\sqrt{s}=1000\,\text{GeV}$, with $s=(p_1+p_2)^2$ and using the physical $Z$ boson mass $m_Z=91.1876\,\text{GeV}$, such that $q_2^2=m_Z^2$.
We set the UV mass parameter to $M=\sqrt{s}$.
Table~\ref{tab:me1} shows the contribution from massless quark-loops, which we verified against the analytic computation of ref.~\cite{Gehrmann:2011ab}.
Table~\ref{tab:me2} displays the contributions from bottom- and top-quark loops, using $m_b = 4.18\,\text{GeV}$ and $m_t = 172\,\text{GeV}$.
\par
In the table headers, $N_p$ denotes the number of Monte Carlo samples, $\Delta\,[\%]$ the relative error in percent and $\Delta\,[\sigma]$ the deviation from the reference value (where available) expressed in units of the Monte Carlo error.
To our knowledge, no benchmark results for these massive contributions are available in the literature.
State-of-the-art analytic results for two-loop (massless) diphoton amplitudes with heavy-quark loops, involving one fewer mass scale than our process, were presented in refs.~\cite{Becchetti:2025rrz,Ahmed:2025osb} and used as a benchmark in our earlier work \cite{Kermanschah:2025wlo}.
\par
Alongside the squared matrix element, the table reports the planar, non-planar, and UV counterterm contributions defined in eq.~\eqref{eq:parts}.
These components were summed and their Monte Carlo errors combined in quadrature, to obtain the squared matrix element.
Since the UV counterterms do not depend on the mass of the quark in the loop, their values are identical in all cases and are shown for completeness.
\par

\input{figures/table_za_heavy}

Because the bottom-quark mass is small, its contribution is numerically close to the massless result, particularly for the planar box diagrams.
Differences are more visible in the non-planar contributions, leading to percent-level effects in the squared matrix element.
In contrast, the large top-quark mass significantly modifies both the analytic structure and the numerical result: only $s$-channel threshold singularities remain as the final-state thresholds are absent.
\par

\input{figures/table_za_ps}

We also evaluate the double-virtual corrections that enter the $pp \to Z\gamma$ cross section, by convoluting the squared matrix element  with the PDFs and integrating over the phase space:
\begin{align}
\label{eq:double-virtual}
    \fulltwoloopfloopscs &= \sum_{q}\int \dd x_1 \dd x_2 f_q(x_1,\mu_F)f_{\bar q}(x_2,\mu_F)v_q^Z v_q^\gamma \frac{F^{(2,\floopsza)}(x_1x_2s)}{2 x_1 x_2 s} \mathcal{O},
    \end{align}
where $f_{q,\bar q}(x_{1,2},\mu_F)$ are the PDFs of the incoming (anti-)quarks at factorization scale $\mu_F$ and momentum fractions $x_{1,2}$, respectively.
By $\dd\Pi_2$ we denote the 2-body phase space measure and $\mathcal{O}$ is an observable function.
We perform the Monte Carlo integration simultaneously over spatial loop momentum, phase space and momentum fractions.
Here, except for $v_q^Z v_q^\gamma$, all other couplings and color factors (as in eqs.~\eqref{eq:order} and \eqref{eq:tree}) are excluded from $\fulltwoloopfloopscs$, which is also not averaged over spin helicities.
\par
The presence of heavy-quark thresholds depends on the center-of-mass energy, and, when integrating over the Bjorken variables, the overlap structure of the threshold surfaces changes.
Compared to ref.~\cite{Kermanschah:2025wlo}, this required minor structural modifications to our pipeline, enabling the dynamic activation of the relevant threshold counterterms and a more flexible choice of their parameterization, so that the double-virtual corrections with heavy-quark loops can be integrated numerically over the full phase space.
\par
In table~\ref{tab:double-virtual}, we show the phase-space integrated double-virtual corrections defined in eq.~\eqref{eq:double-virtual} along with the planar, non-planar and UV counterterm contributions defined according to eq.~\eqref{eq:parts}.
Due to symmetry, the phase-space integrated contributions from the two planar double box integrals are identical, which we denote by $\twoloopfloopsplanarcs = \twoloopfloopsplanaronecs=\twoloopfloopsplanartwocs$ in the table.
We aimed for below 1\% precision on the double-virtual corrections, but significantly higher accuracy was needed for the planar and non-planar contributions due to cancellations\footnote{It would be beneficial to realize these cancellations locally before loop integration, although this has not been investigated further.}.
We also point out the significantly larger number of samples required for the massless quark-loop contributions compared to the massive ones.
\par
We set the center-of-mass energy to $\sqrt{s}=13\,\text{TeV}$ and the factorization scale and UV mass to $\mu_F=M=m_Z=91.1876\,\text{GeV}$.
The observable $\mathcal{O}$ imposes a phase-space cut requiring minimal transverse momenta of $p^\text{min}_T = 50 \,\text{GeV}$ for the photon and $25 \,\text{GeV}$ for the $Z$ boson.
We use the CT10 NLO PDF set \cite{Lai:2010vv} from \textsc{LHAPDF} \cite{Buckley:2014ana} with five light flavors ($u,d,c,s,b$) and convolute it with the squared matrix elements, including massless-quark loops and heavy bottom- and top-quark box loops with $m_b = 4.18\,\text{GeV}$ and $m_t = 172\,\text{GeV}$.
\par

\input{figures/ps_point}

\input{figures/timing}

The numerical integration is carried out using a multi-channel Monte Carlo method with the parameterizations described in ref.~\cite{Kermanschah:2025wlo}.
To ensure numerical stability, each phase-space sample is reevaluated using an equivalent rotated momentum configuration in double precision.
If fewer than seven digits agree, the point is recomputed in double-double precision~\cite{double_double,twofloat_dev}.
All Monte Carlo integrations were performed on the Euler computing cluster at ETH Zurich.
Average evaluation times per sample, benchmarked on a single core of an \texttt{AMD EPYC 7763} processor, are shown in table~\ref{tab:timings}.

\section{Conclusion}
In this work, we applied the numerical techniques and implementation of ref.~\cite{Kermanschah:2025wlo} to compute the two-loop QCD corrections with light- and heavy-quark box loops to $q\bar q \to Z\gamma$.
We validated the massless QCD result by comparing it with the analytic calculation of ref.~\cite{Gehrmann:2011ab} and provided new results are available for the heavy-quark contributions.
\par
We further performed the convolution with PDFs and integrated over the phase space, obtaining the corresponding double-virtual corrections to $Z\gamma$ production in proton–proton collisions at the LHC.
\par
These results highlight the flexibility of our numerical approach, which can accommodate contributions to different processes and treat multiple mass scales within a unified computational framework.
Moreover, the method is capable of providing phase-space integrated corrections with sufficient precision to enable phenomenological studies when combined with real-emission contributions, as available in ref.~\cite{Grazzini:2017mhc,Grazzini:2013bna,Grazzini:2015nwa,Campbell:2017aul}.
\par
We are currently extending this framework to compute the remaining two-loop contributions.

\acknowledgments
We thank the organizers of RADCOR 2025 for the opportunity to present our work and for their hospitality.
We acknowledge the use of the Euler cluster at ETH Zurich for the numerical computations.
This work was supported by the Swiss National Science Foundation through its Postdoc.Mobility funding scheme (grant number 230593) and project funding scheme (grant number 10001706).
\raggedbottom

\newpage
\bibliographystyle{JHEP}
\bibliography{main}

\end{document}

%% file: figures/diags.tex
\begin{figure}[b]
\centering
\begin{subfigure}[t]{0.32\textwidth}
\centering
\begin{tikzpicture}[scale=1.6]
\coordinate (a) at (1.5,1);
\coordinate (b) at (2.5,1);
\coordinate (c) at (3,0.5);
\coordinate (d) at (2.5,0);
\coordinate (e) at (1.5,0);
\draw[cut,orange] (a) -- 
(e);
\draw[cut,gray] (b) -- 
(d);
\draw[cut,violet] (d) -- 
(c);
\draw[cut,cyan] (e) -- 
(c);
\draw[cut,magenta] (a) -- 
(d);
\draw[cut,teal] (e) -- 
(b);
\begin{feynman}
\vertex (A) at (0.5,1);
\vertex (B) at (1,1);
\vertex (C) at (2,1);
\vertex (D) at (3,1);
\vertex (E) at (3.5,1);
\vertex (F) at (0.5,0);
\vertex (G) at (1,0);
\vertex (H) at (2,0);
\vertex (I) at (3,0);
\vertex (J) at (3.5,0);
\diagram*{
(A) -- [fermion, momentum=$p_1$] (B),
(I) -- [boson, momentum'=$q_2$] (J),
(G) -- [fermion, reversed momentum=$p_2$] (F),
(D) -- [boson, momentum=$q_1$] (E),
(B) -- [gluon] (C),
(G) -- [gluon] (H),
(B) -- [fermion, reversed momentum'=$k_1$] (G),
(D) -- [fermion, very thick, momentum'=$k_2$] (C),
(H) -- [fermion, very thick] (I),
(C) -- [fermion, very thick] (H),
(I) -- [fermion, very thick] (D)
};
\end{feynman}
\end{tikzpicture}
\caption{$G_{\planar_1}$}
\end{subfigure}
\hfill
\begin{subfigure}[t]{0.32\textwidth}
\centering
\begin{tikzpicture}[scale=1.6]
\coordinate (a) at (1.5,1);
\coordinate (b) at (2.5,1);
\coordinate (c) at (3,0.5);
\coordinate (d) at (2.5,0);
\coordinate (e) at (1.5,0);
\draw[cut,orange] (a) -- 
(e);
\draw[cut,gray] (b) -- 
(d);
\draw[cut,blue] (a) -- 
(c); 
\draw[cut,green] (b) -- 
(c);
\draw[cut,magenta] (a) -- 
(d); 
\draw[cut,teal] (e) -- 
(b);
\begin{feynman}
\vertex (A) at (0.5,1);
\vertex (B) at (1,1);
\vertex (C) at (2,1);
\vertex (D) at (3,1);
\vertex (E) at (3.5,1);
\vertex (F) at (0.5,0);
\vertex (G) at (1,0);
\vertex (H) at (2,0);
\vertex (I) at (3,0);
\vertex (J) at (3.5,0);
\diagram*{
(A) -- [fermion, momentum=$p_1$] (B),
(I) -- [boson, momentum'=$q_1$] (J),
(G) -- [fermion, reversed momentum=$p_2$] (F),
(D) -- [boson, momentum=$q_2$] (E),
(B) -- [gluon] (C),
(G) -- [gluon] (H),
(B) -- [fermion, reversed momentum'=$k_1$] (G),
(D) -- [fermion, very thick, momentum'=$k_2$] (C),
(H) -- [fermion, very thick] (I),
(C) -- [fermion, very thick] (H),
(I) -- [fermion, very thick] (D)
};
\end{feynman}
\end{tikzpicture}
\caption{$G_{\planar_2}$}
\end{subfigure}
\hfill
\begin{subfigure}[t]{0.32\textwidth}
\centering
\begin{tikzpicture}[scale=1.6]
\draw[cut,orange] (1.25,1) -- 
(1.25,0); 
\draw[cut,blue]  (2,1) 
-- (2.75,0.25);
\draw[cut,green] (2.75,0.75) -- 
(2.75,0.25);
\draw[cut,teal] (1.75,0) .. controls (1.75,0.75) .. 
(2.75,0.75); 
\draw[cut,purple] (1.75,1) .. controls (1.75,0.25) .. 
(2.75,0.25);
\draw[cut,pink] (2,0) 
-- (2.75,0.75);
\begin{feynman}
\vertex (A) at (0.5,1);
\vertex (B) at (1,1);
\vertex (C) at (2.5,1);
\vertex (D) at (3,0.5);
\vertex (E) at (3.5,0.5);
\vertex (F) at (0.5,0);
\vertex (G) at (1,0);
\vertex (H) at (2,0.5);
\vertex (I) at (2.5,0);
\vertex (J) at (1.5,0.5);
\diagram*{
(A) -- [fermion, momentum=$p_1$] (B),
(H) -- [boson, momentum=$q_1$] (J),
(G) -- [fermion, reversed momentum=$p_2$] (F),
(D) -- [boson, momentum'=$q_2$] (E),
(B) -- [gluon] (C),
(I) -- [gluon] (G),
(B) -- [fermion, reversed momentum'=$k_1$] (G),
(D) -- [fermion, very thick, momentum'=$k_2$] (C),
(H) -- [fermion, very thick] (I),
(C) -- [fermion, very thick] (H),
(I) -- [fermion, very thick] (D)
};
\end{feynman}
\end{tikzpicture}
\caption{$G_{\nonplanar}$}
\end{subfigure}
\caption{Planar and non-planar double box Feynman diagrams and their Cutkosky cuts identifying the threshold singularities. The diagrams with reversed charge flow are not shown.}
\label{fig:diagrams}
\end{figure}
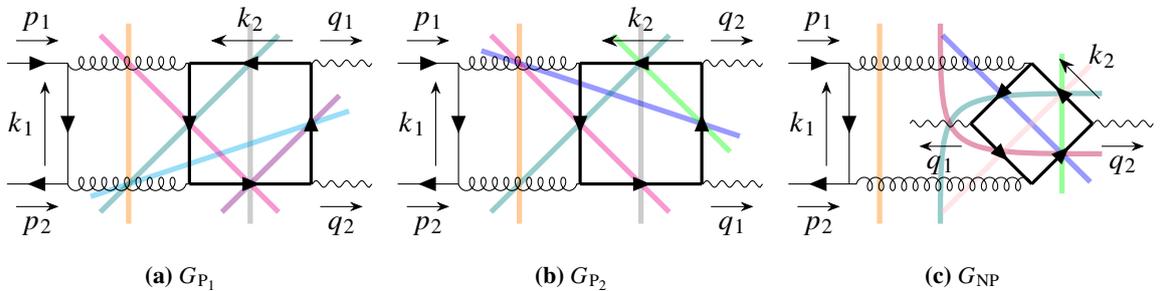

%% file: figures/table_za_light.tex
\begin{table}[t]
\centering
\renewcommand{\arraystretch}{1.28}
\resizebox{\columnwidth}
{!}{%
\begin{tabular}{lllrrccrr}
\hline\hline
Process & PSP & Part & $N_p \ [10^8]$ & Exp. & Reference & Result & $\Delta \ [\sigma]$ & $\Delta \ [\%]$ \\
\hline\hline\multirow{15}{*}{\makecell[l]{$q\bar{q}\to Z\gamma$\\$m=0$}}
& \multirow{5}{*}{1}
& \multirow{1}{*}{$\twoloopfloopsplanarone$}
& {\ttlf{$\texttt{5.2}$}}
& {\ttlf$\texttt{10}^{\texttt{+2}}$}
&
& {\ttlf\texttt{-3.2119~$\pmtt$~0.0058}}
&
& {\ttlf\texttt{0.181}}
\\
\cline{3-3}\cline{4-4}\cline{5-5}\cline{6-6}\cline{7-7}\cline{8-8}\cline{9-9}& & \multirow{1}{*}{$\twoloopfloopsplanartwo$}
& {\ttlf{$\texttt{5.2}$}}
& {\ttlf$\texttt{10}^{\texttt{+2}}$}
&
& {\ttlf\texttt{-3.5942~$\pmtt$~0.0047}}
&
& {\ttlf\texttt{0.132}}
\\
\cline{3-3}\cline{4-4}\cline{5-5}\cline{6-6}\cline{7-7}\cline{8-8}\cline{9-9}& & \multirow{1}{*}{$\twoloopfloopsnonplanar$}
& {\ttlf{$\texttt{5.2}$}}
& {\ttlf$\texttt{10}^{\texttt{+2}}$}
&
& {\ttlf\texttt{~4.2234~$\pmtt$~0.0059}}
&
& {\ttlf\texttt{0.141}}
\\
\cline{3-3}\cline{4-4}\cline{5-5}\cline{6-6}\cline{7-7}\cline{8-8}\cline{9-9}& & \multirow{1}{*}{$\twoloopfloopsuv$}
& {\ttlf{$\texttt{5.2}$}}
& {\ttlf$\texttt{10}^{\texttt{+1}}$}
&
& {\ttlf\texttt{-6.4735~$\pmtt$~0.0010}}
&
& {\ttlf\texttt{0.016}}
\\
\ccline{3-3}\ccline{4-4}\ccline{5-5}\ccline{6-6}\ccline{7-7}\ccline{8-8}\ccline{9-9}& & \multirow{1}{*}{$\boldsymbol\fulltwoloopfloops$}
&
& {\ttbf$\texttt{10}^{\texttt{+2}}$}
& {\ttbf\texttt{-5.8082}}
& {\ttbf\texttt{-5.8128~$\pmtt$~0.0191}}
& {\ttbf\texttt{0.240}}
& {\ttbf\texttt{0.329}}
\\
[\doublerulesep]\ccline{2-2}\ccline{3-3}\ccline{4-4}\ccline{5-5}\ccline{6-6}\ccline{7-7}\ccline{8-8}\ccline{9-9}& \multirow{5}{*}{2}
& \multirow{1}{*}{$\twoloopfloopsplanarone$}
& {\ttlf{$\texttt{5.2}$}}
& {\ttlf$\texttt{10}^{\texttt{+2}}$}
&
& {\ttlf\texttt{-3.5100~$\pmtt$~0.0042}}
&
& {\ttlf\texttt{0.121}}
\\
\cline{3-3}\cline{4-4}\cline{5-5}\cline{6-6}\cline{7-7}\cline{8-8}\cline{9-9}& & \multirow{1}{*}{$\twoloopfloopsplanartwo$}
& {\ttlf{$\texttt{5.2}$}}
& {\ttlf$\texttt{10}^{\texttt{+2}}$}
&
& {\ttlf\texttt{-3.2959~$\pmtt$~0.0060}}
&
& {\ttlf\texttt{0.182}}
\\
\cline{3-3}\cline{4-4}\cline{5-5}\cline{6-6}\cline{7-7}\cline{8-8}\cline{9-9}& & \multirow{1}{*}{$\twoloopfloopsnonplanar$}
& {\ttlf{$\texttt{5.2}$}}
& {\ttlf$\texttt{10}^{\texttt{+2}}$}
&
& {\ttlf\texttt{~4.3416~$\pmtt$~0.0051}}
&
& {\ttlf\texttt{0.118}}
\\
\cline{3-3}\cline{4-4}\cline{5-5}\cline{6-6}\cline{7-7}\cline{8-8}\cline{9-9}& & \multirow{1}{*}{$\twoloopfloopsuv$}
& {\ttlf{$\texttt{5.2}$}}
& {\ttlf$\texttt{10}^{\texttt{+1}}$}
&
& {\ttlf\texttt{-6.4727~$\pmtt$~0.0008}}
&
& {\ttlf\texttt{0.013}}
\\
\ccline{3-3}\ccline{4-4}\ccline{5-5}\ccline{6-6}\ccline{7-7}\ccline{8-8}\ccline{9-9}& & \multirow{1}{*}{$\boldsymbol\fulltwoloopfloops$}
&
& {\ttbf$\texttt{10}^{\texttt{+2}}$}
& {\ttbf\texttt{-5.5645}}
& {\ttbf\texttt{-5.5758~$\pmtt$~0.0179}}
& {\ttbf\texttt{0.630}}
& {\ttbf\texttt{0.322}}
\\
[\doublerulesep]\ccline{2-2}\ccline{3-3}\ccline{4-4}\ccline{5-5}\ccline{6-6}\ccline{7-7}\ccline{8-8}\ccline{9-9}& \multirow{5}{*}{3}
& \multirow{1}{*}{$\twoloopfloopsplanarone$}
& {\ttlf{$\texttt{8.2}$}}
& {\ttlf$\texttt{10}^{\texttt{+2}}$}
&
& {\ttlf\texttt{-2.6861~$\pmtt$~0.0078}}
&
& {\ttlf\texttt{0.292}}
\\
\cline{3-3}\cline{4-4}\cline{5-5}\cline{6-6}\cline{7-7}\cline{8-8}\cline{9-9}& & \multirow{1}{*}{$\twoloopfloopsplanartwo$}
& {\ttlf{$\texttt{5.2}$}}
& {\ttlf$\texttt{10}^{\texttt{+2}}$}
&
& {\ttlf\texttt{-3.8378~$\pmtt$~0.0063}}
&
& {\ttlf\texttt{0.164}}
\\
\cline{3-3}\cline{4-4}\cline{5-5}\cline{6-6}\cline{7-7}\cline{8-8}\cline{9-9}& & \multirow{1}{*}{$\twoloopfloopsnonplanar$}
& {\ttlf{$\texttt{5.2}$}}
& {\ttlf$\texttt{10}^{\texttt{+2}}$}
&
& {\ttlf\texttt{~3.7536~$\pmtt$~0.0096}}
&
& {\ttlf\texttt{0.256}}
\\
\cline{3-3}\cline{4-4}\cline{5-5}\cline{6-6}\cline{7-7}\cline{8-8}\cline{9-9}& & \multirow{1}{*}{$\twoloopfloopsuv$}
& {\ttlf{$\texttt{5.2}$}}
& {\ttlf$\texttt{10}^{\texttt{+1}}$}
&
& {\ttlf\texttt{-6.4738~$\pmtt$~0.0017}}
&
& {\ttlf\texttt{0.026}}
\\
\ccline{3-3}\ccline{4-4}\ccline{5-5}\ccline{6-6}\ccline{7-7}\ccline{8-8}\ccline{9-9}& & \multirow{1}{*}{$\boldsymbol\fulltwoloopfloops$}
&
& {\ttbf$\texttt{10}^{\texttt{+2}}$}
& {\ttbf\texttt{-6.2120}}
& {\ttbf\texttt{-6.1879~$\pmtt$~0.0278}}
& {\ttbf\texttt{0.866}}
& {\ttbf\texttt{0.450}}
\\
\cline{4-4}\cline{5-5}\cline{6-6}\cline{7-7}\cline{8-8}\cline{9-9}\hline\hline
\end{tabular}%
}%
\caption{\label{tab:me1}
Helicity-summed squared matrix element (in bold), defined in eq.~\eqref{eq:matrix-element}, together with the planar, non-planar, and UV counterterm contributions introduced in eq.~\eqref{eq:parts}, evaluated at the three phase-space points given in listing~\ref{lst:ps_points}, for the two-loop corrections involving massless-quark box loops.
The reference results were obtained from the analytic amplitudes computed in ref.~\cite{Gehrmann:2011ab}.
}
\end{table}

%% file: figures/table_za_heavy.tex
\begin{table}[H]
\centering
\renewcommand{\arraystretch}{1.28}
\begin{tabular}{lllrrcr}
\hline\hline
Process & PSP & Part & $N_p \ [10^8]$ & Exp. & Result & $\Delta \ [\%]$ \\
\hline\hline\multirow{15}{*}{\makecell[l]{$q\bar{q}\to Z\gamma$\\$m=m_b$}}
& \multirow{5}{*}{1}
& \multirow{1}{*}{$\twoloopfloopsplanarone$}
& {\ttlf{$\texttt{5.2}$}}
& {\ttlf$\texttt{10}^{\texttt{+2}}$}
& {\ttlf\texttt{-3.2121~$\pmtt$~0.0055}}
& {\ttlf\texttt{0.171}}
\\
\cline{3-3}\cline{4-4}\cline{5-5}\cline{6-6}\cline{7-7}& & \multirow{1}{*}{$\twoloopfloopsplanartwo$}
& {\ttlf{$\texttt{5.2}$}}
& {\ttlf$\texttt{10}^{\texttt{+2}}$}
& {\ttlf\texttt{-3.5921~$\pmtt$~0.0037}}
& {\ttlf\texttt{0.104}}
\\
\cline{3-3}\cline{4-4}\cline{5-5}\cline{6-6}\cline{7-7}& & \multirow{1}{*}{$\twoloopfloopsnonplanar$}
& {\ttlf{$\texttt{5.2}$}}
& {\ttlf$\texttt{10}^{\texttt{+2}}$}
& {\ttlf\texttt{~4.1129~$\pmtt$~0.0043}}
& {\ttlf\texttt{0.104}}
\\
\cline{3-3}\cline{4-4}\cline{5-5}\cline{6-6}\cline{7-7}& & \multirow{1}{*}{$\twoloopfloopsuv$}
& {\ttlf{$\texttt{5.2}$}}
& {\ttlf$\texttt{10}^{\texttt{+1}}$}
& {\ttlf\texttt{-6.4735~$\pmtt$~0.0010}}
& {\ttlf\texttt{0.016}}
\\
\ccline{3-3}\ccline{4-4}\ccline{5-5}\ccline{6-6}\ccline{7-7}& & \multirow{1}{*}{$\boldsymbol\fulltwoloopfloops$}
&
& {\ttbf$\texttt{10}^{\texttt{+2}}$}
& {\ttbf\texttt{-6.0300~$\pmtt$~0.0158}}
& {\ttbf\texttt{0.262}}
\\
[\doublerulesep]\ccline{2-2}\ccline{3-3}\ccline{4-4}\ccline{5-5}\ccline{6-6}\ccline{7-7}& \multirow{5}{*}{2}
& \multirow{1}{*}{$\twoloopfloopsplanarone$}
& {\ttlf{$\texttt{5.2}$}}
& {\ttlf$\texttt{10}^{\texttt{+2}}$}
& {\ttlf\texttt{-3.5063~$\pmtt$~0.0037}}
& {\ttlf\texttt{0.105}}
\\
\cline{3-3}\cline{4-4}\cline{5-5}\cline{6-6}\cline{7-7}& & \multirow{1}{*}{$\twoloopfloopsplanartwo$}
& {\ttlf{$\texttt{5.2}$}}
& {\ttlf$\texttt{10}^{\texttt{+2}}$}
& {\ttlf\texttt{-3.3003~$\pmtt$~0.0047}}
& {\ttlf\texttt{0.142}}
\\
\cline{3-3}\cline{4-4}\cline{5-5}\cline{6-6}\cline{7-7}& & \multirow{1}{*}{$\twoloopfloopsnonplanar$}
& {\ttlf{$\texttt{5.2}$}}
& {\ttlf$\texttt{10}^{\texttt{+2}}$}
& {\ttlf\texttt{~4.2485~$\pmtt$~0.0039}}
& {\ttlf\texttt{0.092}}
\\
\cline{3-3}\cline{4-4}\cline{5-5}\cline{6-6}\cline{7-7}& & \multirow{1}{*}{$\twoloopfloopsuv$}
& {\ttlf{$\texttt{5.2}$}}
& {\ttlf$\texttt{10}^{\texttt{+1}}$}
& {\ttlf\texttt{-6.4727~$\pmtt$~0.0008}}
& {\ttlf\texttt{0.013}}
\\
\ccline{3-3}\ccline{4-4}\ccline{5-5}\ccline{6-6}\ccline{7-7}& & \multirow{1}{*}{$\boldsymbol\fulltwoloopfloops$}
&
& {\ttbf$\texttt{10}^{\texttt{+2}}$}
& {\ttbf\texttt{-5.7634~$\pmtt$~0.0142}}
& {\ttbf\texttt{0.247}}
\\
[\doublerulesep]\ccline{2-2}\ccline{3-3}\ccline{4-4}\ccline{5-5}\ccline{6-6}\ccline{7-7}& \multirow{5}{*}{3}
& \multirow{1}{*}{$\twoloopfloopsplanarone$}
& {\ttlf{$\texttt{5.8}$}}
& {\ttlf$\texttt{10}^{\texttt{+2}}$}
& {\ttlf\texttt{-2.7072~$\pmtt$~0.0081}}
& {\ttlf\texttt{0.299}}
\\
\cline{3-3}\cline{4-4}\cline{5-5}\cline{6-6}\cline{7-7}& & \multirow{1}{*}{$\twoloopfloopsplanartwo$}
& {\ttlf{$\texttt{5.2}$}}
& {\ttlf$\texttt{10}^{\texttt{+2}}$}
& {\ttlf\texttt{-3.8490~$\pmtt$~0.0047}}
& {\ttlf\texttt{0.123}}
\\
\cline{3-3}\cline{4-4}\cline{5-5}\cline{6-6}\cline{7-7}& & \multirow{1}{*}{$\twoloopfloopsnonplanar$}
& {\ttlf{$\texttt{5.2}$}}
& {\ttlf$\texttt{10}^{\texttt{+2}}$}
& {\ttlf\texttt{~3.5838~$\pmtt$~0.0059}}
& {\ttlf\texttt{0.166}}
\\
\cline{3-3}\cline{4-4}\cline{5-5}\cline{6-6}\cline{7-7}& & \multirow{1}{*}{$\twoloopfloopsuv$}
& {\ttlf{$\texttt{5.2}$}}
& {\ttlf$\texttt{10}^{\texttt{+1}}$}
& {\ttlf\texttt{-6.4738~$\pmtt$~0.0017}}
& {\ttlf\texttt{0.026}}
\\
\ccline{3-3}\ccline{4-4}\ccline{5-5}\ccline{6-6}\ccline{7-7}& & \multirow{1}{*}{$\boldsymbol\fulltwoloopfloops$}
&
& {\ttbf$\texttt{10}^{\texttt{+2}}$}
& {\ttbf\texttt{-6.5922~$\pmtt$~0.0222}}
& {\ttbf\texttt{0.337}}
\\
[\doublerulesep]\ccline{1-1}\ccline{2-2}\ccline{3-3}\ccline{4-4}\ccline{5-5}\ccline{6-6}\ccline{7-7}\multirow{15}{*}{\makecell[l]{$q\bar{q}\to Z\gamma$\\$m=m_t$}}
& \multirow{5}{*}{1}
& \multirow{1}{*}{$\twoloopfloopsplanarone$}
& {\ttlf{$\texttt{5.2}$}}
& {\ttlf$\texttt{10}^{\texttt{+2}}$}
& {\ttlf\texttt{-5.3468~$\pmtt$~0.0022}}
& {\ttlf\texttt{0.042}}
\\
\cline{3-3}\cline{4-4}\cline{5-5}\cline{6-6}\cline{7-7}& & \multirow{1}{*}{$\twoloopfloopsplanartwo$}
& {\ttlf{$\texttt{5.2}$}}
& {\ttlf$\texttt{10}^{\texttt{+2}}$}
& {\ttlf\texttt{-4.5173~$\pmtt$~0.0016}}
& {\ttlf\texttt{0.035}}
\\
\cline{3-3}\cline{4-4}\cline{5-5}\cline{6-6}\cline{7-7}& & \multirow{1}{*}{$\twoloopfloopsnonplanar$}
& {\ttlf{$\texttt{5.2}$}}
& {\ttlf$\texttt{10}^{\texttt{+2}}$}
& {\ttlf\texttt{~9.3880~$\pmtt$~0.0018}}
& {\ttlf\texttt{0.019}}
\\
\cline{3-3}\cline{4-4}\cline{5-5}\cline{6-6}\cline{7-7}& & \multirow{1}{*}{$\twoloopfloopsuv$}
& {\ttlf{$\texttt{5.2}$}}
& {\ttlf$\texttt{10}^{\texttt{+1}}$}
& {\ttlf\texttt{-6.4735~$\pmtt$~0.0010}}
& {\ttlf\texttt{0.016}}
\\
\ccline{3-3}\ccline{4-4}\ccline{5-5}\ccline{6-6}\ccline{7-7}& & \multirow{1}{*}{$\boldsymbol\fulltwoloopfloops$}
&
& {\ttbf$\texttt{10}^{\texttt{+2}}$}
& {\ttbf\texttt{-1.5996~$\pmtt$~0.0065}}
& {\ttbf\texttt{0.407}}
\\
[\doublerulesep]\ccline{2-2}\ccline{3-3}\ccline{4-4}\ccline{5-5}\ccline{6-6}\ccline{7-7}& \multirow{5}{*}{2}
& \multirow{1}{*}{$\twoloopfloopsplanarone$}
& {\ttlf{$\texttt{5.2}$}}
& {\ttlf$\texttt{10}^{\texttt{+2}}$}
& {\ttlf\texttt{-4.4679~$\pmtt$~0.0016}}
& {\ttlf\texttt{0.035}}
\\
\cline{3-3}\cline{4-4}\cline{5-5}\cline{6-6}\cline{7-7}& & \multirow{1}{*}{$\twoloopfloopsplanartwo$}
& {\ttlf{$\texttt{5.2}$}}
& {\ttlf$\texttt{10}^{\texttt{+2}}$}
& {\ttlf\texttt{-4.9993~$\pmtt$~0.0021}}
& {\ttlf\texttt{0.042}}
\\
\cline{3-3}\cline{4-4}\cline{5-5}\cline{6-6}\cline{7-7}& & \multirow{1}{*}{$\twoloopfloopsnonplanar$}
& {\ttlf{$\texttt{5.2}$}}
& {\ttlf$\texttt{10}^{\texttt{+2}}$}
& {\ttlf\texttt{~9.0230~$\pmtt$~0.0016}}
& {\ttlf\texttt{0.017}}
\\
\cline{3-3}\cline{4-4}\cline{5-5}\cline{6-6}\cline{7-7}& & \multirow{1}{*}{$\twoloopfloopsuv$}
& {\ttlf{$\texttt{5.2}$}}
& {\ttlf$\texttt{10}^{\texttt{+1}}$}
& {\ttlf\texttt{-6.4727~$\pmtt$~0.0008}}
& {\ttlf\texttt{0.013}}
\\
\ccline{3-3}\ccline{4-4}\ccline{5-5}\ccline{6-6}\ccline{7-7}& & \multirow{1}{*}{$\boldsymbol\fulltwoloopfloops$}
&
& {\ttbf$\texttt{10}^{\texttt{+2}}$}
& {\ttbf\texttt{-1.5356~$\pmtt$~0.0061}}
& {\ttbf\texttt{0.399}}
\\
[\doublerulesep]\ccline{2-2}\ccline{3-3}\ccline{4-4}\ccline{5-5}\ccline{6-6}\ccline{7-7}& \multirow{5}{*}{3}
& \multirow{1}{*}{$\twoloopfloopsplanarone$}
& {\ttlf{$\texttt{5.2}$}}
& {\ttlf$\texttt{10}^{\texttt{+2}}$}
& {\ttlf\texttt{-6.4309~$\pmtt$~0.0031}}
& {\ttlf\texttt{0.049}}
\\
\cline{3-3}\cline{4-4}\cline{5-5}\cline{6-6}\cline{7-7}& & \multirow{1}{*}{$\twoloopfloopsplanartwo$}
& {\ttlf{$\texttt{5.2}$}}
& {\ttlf$\texttt{10}^{\texttt{+2}}$}
& {\ttlf\texttt{-4.8192~$\pmtt$~0.0018}}
& {\ttlf\texttt{0.037}}
\\
\cline{3-3}\cline{4-4}\cline{5-5}\cline{6-6}\cline{7-7}& & \multirow{1}{*}{$\twoloopfloopsnonplanar$}
& {\ttlf{$\texttt{5.2}$}}
& {\ttlf$\texttt{10}^{\texttt{+3}}$}
& {\ttlf\texttt{~1.0681~$\pmtt$~0.0003}}
& {\ttlf\texttt{0.024}}
\\
\cline{3-3}\cline{4-4}\cline{5-5}\cline{6-6}\cline{7-7}& & \multirow{1}{*}{$\twoloopfloopsuv$}
& {\ttlf{$\texttt{5.2}$}}
& {\ttlf$\texttt{10}^{\texttt{+1}}$}
& {\ttlf\texttt{-6.4738~$\pmtt$~0.0017}}
& {\ttlf\texttt{0.026}}
\\
\ccline{3-3}\ccline{4-4}\ccline{5-5}\ccline{6-6}\ccline{7-7}& & \multirow{1}{*}{$\boldsymbol\fulltwoloopfloops$}
&
& {\ttbf$\texttt{10}^{\texttt{+2}}$}
& {\ttbf\texttt{-1.7862~$\pmtt$~0.0089}}
& {\ttbf\texttt{0.497}}
\\
\cline{4-4}\cline{5-5}\cline{6-6}\cline{7-7}\hline\hline
\end{tabular}%
\caption{\label{tab:me2}
Helicity-summed squared matrix element (in bold), defined in eq.~\eqref{eq:matrix-element}, together with the planar, non-planar, and UV counterterm contributions introduced in eq.~\eqref{eq:parts}, evaluated at the three phase-space points given in listing~\ref{lst:ps_points}, for the two-loop corrections with top- and bottom-quark box loops.
}%
\end{table}

%% file: figures/table_za_ps.tex
\begin{table}[t]
\centering
\renewcommand{\arraystretch}{1.4}
\begin{tabular}{llrrcr}
\hline\hline
Process & Part & $N_p \ [10^8]$ & Exp. & Result & $\Delta \ [\%]$ \\
\hline\hline\multirow{4}{*}{\makecell[l]{$pp\to Z\gamma$\\$m=0$}}
& $\twoloopfloopsplanarcs$
& \ttlf{$\texttt{262.9}$}
& {\ttlf$\texttt{10}^{\texttt{-5}}$}
& {\ttlf\texttt{~8.0566~$\pmtt$~0.0028}}
& {\ttlf\texttt{0.035}}
\\
\cline{2-2}\cline{3-3}\cline{4-4}\cline{5-5}\cline{6-6}& $\twoloopfloopsnonplanarcs$
& \ttlf{$\texttt{108.3}$}
& {\ttlf$\texttt{10}^{\texttt{-4}}$}
& {\ttlf\texttt{-1.6398~$\pmtt$~0.0006}}
& {\ttlf\texttt{0.035}}
\\
\cline{2-2}\cline{3-3}\cline{4-4}\cline{5-5}\cline{6-6}& $\twoloopfloopsuvcs$
& \ttlf{$\texttt{5.2}$}
& {\ttlf$\texttt{10}^{\texttt{-5}}$}
& {\ttlf\texttt{~2.2703~$\pmtt$~0.0006}}
& {\ttlf\texttt{0.027}}
\\
\ccline{2-2}\ccline{3-3}\ccline{4-4}\ccline{5-5}\ccline{6-6}& $\boldsymbol\fulltwoloopfloopscs$
&
& {\ttbf$\texttt{10}^{\texttt{-5}}$}
& {\ttbf\texttt{~1.7005~$\pmtt$~0.0161}}
& {\ttbf\texttt{0.945}}
\\
\ccline{1-1}\ccline{2-2}\ccline{3-3}\ccline{4-4}\ccline{5-5}\ccline{6-6}\multirow{4}{*}{\makecell[l]{$pp\to Z\gamma$\\$m=m_b$}}
& $\twoloopfloopsplanarcs$
& \ttlf{$\texttt{37.0}$}
& {\ttlf$\texttt{10}^{\texttt{-5}}$}
& {\ttlf\texttt{~8.1366~$\pmtt$~0.0071}}
& {\ttlf\texttt{0.087}}
\\
\cline{2-2}\cline{3-3}\cline{4-4}\cline{5-5}\cline{6-6}& $\twoloopfloopsnonplanarcs$
& \ttlf{$\texttt{17.9}$}
& {\ttlf$\texttt{10}^{\texttt{-4}}$}
& {\ttlf\texttt{-1.5434~$\pmtt$~0.0009}}
& {\ttlf\texttt{0.058}}
\\
\cline{2-2}\cline{3-3}\cline{4-4}\cline{5-5}\cline{6-6}& $\twoloopfloopsuvcs$
& \ttlf{$\texttt{5.2}$}
& {\ttlf$\texttt{10}^{\texttt{-5}}$}
& {\ttlf\texttt{~2.2703~$\pmtt$~0.0006}}
& {\ttlf\texttt{0.027}}
\\
\ccline{2-2}\ccline{3-3}\ccline{4-4}\ccline{5-5}\ccline{6-6}& $\boldsymbol\fulltwoloopfloopscs$
&
& {\ttbf$\texttt{10}^{\texttt{-5}}$}
& {\ttbf\texttt{~3.9486~$\pmtt$~0.0334}}
& {\ttbf\texttt{0.846}}
\\
\ccline{1-1}\ccline{2-2}\ccline{3-3}\ccline{4-4}\ccline{5-5}\ccline{6-6}\multirow{4}{*}{\makecell[l]{$pp\to Z\gamma$\\$m=m_t$}}
& $\twoloopfloopsplanarcs$
& \ttlf{$\texttt{33.6}$}
& {\ttlf$\texttt{10}^{\texttt{-5}}$}
& {\ttlf\texttt{~3.5718~$\pmtt$~0.0012}}
& {\ttlf\texttt{0.035}}
\\
\cline{2-2}\cline{3-3}\cline{4-4}\cline{5-5}\cline{6-6}& $\twoloopfloopsnonplanarcs$
& \ttlf{$\texttt{10.4}$}
& {\ttlf$\texttt{10}^{\texttt{-5}}$}
& {\ttlf\texttt{-7.8350~$\pmtt$~0.0027}}
& {\ttlf\texttt{0.034}}
\\
\cline{2-2}\cline{3-3}\cline{4-4}\cline{5-5}\cline{6-6}& $\twoloopfloopsuvcs$
& \ttlf{$\texttt{5.2}$}
& {\ttlf$\texttt{10}^{\texttt{-5}}$}
& {\ttlf\texttt{~2.2703~$\pmtt$~0.0006}}
& {\ttlf\texttt{0.027}}
\\
\ccline{2-2}\ccline{3-3}\ccline{4-4}\ccline{5-5}\ccline{6-6}& $\boldsymbol\fulltwoloopfloopscs$
&
& {\ttbf$\texttt{10}^{\texttt{-6}}$}
& {\ttbf\texttt{~8.8753~$\pmtt$~0.0735}}
& {\ttbf\texttt{0.828}}
\\
\cline{2-2}\cline{3-3}\cline{4-4}\cline{5-5}\cline{6-6}\hline\hline
\end{tabular}%
\caption{\label{tab:double-virtual}Double-virtual corrections as defined in eq.~\eqref{eq:double-virtual} mediated by light- and heavy-quark loops, including the planar, non-planar and UV counterterms contributions defined through eq.~\eqref{eq:parts}.
Collision energy, transverse momentum cuts, and PDF specifications are given in the text.
}%

\end{table}

%% file: figures/ps_point.tex
\begin{figure}[t]
\begin{lstlisting}[language=Python,frame=single,caption={Three phase-space points for $q(p_1) \bar{q}(p_2) \to \gamma(q_1) Z(q_2)$ with $m_Z=91.1876\,\text{GeV}$ at a center-of-mass energy of $\sqrt{s}=1000\,\text{GeV}$.},commentstyle=\color{teal},captionpos=b,label={lst:ps_points}]
# PSP1
0.5000000000000000E+03  0.0000000000000000E+00  0.0000000000000000E+00  0.5000000000000000E+03
0.5000000000000000E+03  0.0000000000000000E+00  0.0000000000000000E+00 -0.5000000000000000E+03
0.4958424108776996E+03  0.1100019292470276E+03  0.4411319421848414E+03 -0.1978936117492713E+03
0.5041575891223005E+03 -0.1100019292470276E+03 -0.4411319421848414E+03  0.1978936117492712E+03
# PSP2
0.5000000000000000E+03  0.0000000000000000E+00  0.0000000000000000E+00  0.5000000000000000E+03
0.5000000000000000E+03  0.0000000000000000E+00  0.0000000000000000E+00 -0.5000000000000000E+03
0.4958424108776997E+03 -0.2795713281286689E+03 -0.3875414700574890E+03  0.1323298072962480E+03
0.5041575891223005E+03  0.2795713281286690E+03  0.3875414700574891E+03 -0.1323298072962480E+03
# PSP3
0.5000000000000000E+03  0.0000000000000000E+00  0.0000000000000000E+00  0.5000000000000000E+03
0.5000000000000000E+03  0.0000000000000000E+00  0.0000000000000000E+00 -0.5000000000000000E+03
0.4958424108776987E+03 -0.2386202223474536E+03  0.2781398312686913E+03 -0.3340034732958395E+03
0.5041575891223014E+03  0.2386202223474535E+03 -0.2781398312686913E+03  0.3340034732958395E+03
\end{lstlisting}
\end{figure}

%% file: figures/timing.tex
\begin{table}[b]
\centering
\renewcommand{\arraystretch}{1.35}
\begin{tabular}{lcccc}
\hline\hline
Process & $\twoloopfloopsplanarone$ & $\twoloopfloopsplanartwo$ & $\twoloopfloopsnonplanar$ & $\twoloopfloopsuv$ \\
\hline\hline
\makecell[l]{$q\bar{q}\to Z\gamma$, $m=0$}
& \ttlf\texttt{0.646} & \ttlf\texttt{0.627} & \ttlf\texttt{0.674} & \ttlf\texttt{0.022}\\
\hline
\makecell[l]{$q\bar{q}\to Z\gamma$, $m=m_b$}
& \ttlf\texttt{1.807} & \ttlf\texttt{1.423} & \ttlf\texttt{1.917} & \ttlf\texttt{0.022} \\
\hline
\makecell[l]{$q\bar{q}\to Z\gamma$, $m=m_t$}
& \ttlf\texttt{1.632} & \ttlf\texttt{1.293} & \ttlf\texttt{1.394} & \ttlf\texttt{0.022}\\
\hline\hline
\end{tabular}%
\caption{\label{tab:timings}Single-core integrand evaluation times in milliseconds using double-precision arithmetic averaged over $10^4$ samples.
In contrast to ref.~\cite{Kermanschah:2025wlo}, expression optimization was performed only at the level of individual diagrams rather than for their combined expression, leading to slightly longer evaluation times.
}
\end{table}

%% file: main.bib
@article{Kermanschah:2024utt,
    author = "Kermanschah, Dario and Vicini, Matilde",
    title = "{N$_{f}$-contribution to the virtual correction for electroweak vector boson production at NNLO}",
    eprint = "2407.18051",
    archivePrefix = "arXiv",
    primaryClass = "hep-ph",
    doi = "10.1007/JHEP09(2025)213",
    journal = "JHEP",
    volume = "09",
    pages = "213",
    year = "2025"
}

@article{Kermanschah:2025wlo,
    author = "Kermanschah, Dario and Vicini, Matilde",
    title = "{Two-loop QCD corrections for real and off-shell diphoton and triphoton production via quark loops}",
    eprint = "2510.18801",
    archivePrefix = "arXiv",
    primaryClass = "hep-ph",
    reportNumber = "OUTP-25-04P",
    month = "10",
    year = "2025"
}

@article{Gehrmann:2011ab,
    author = "Gehrmann, Thomas and Tancredi, Lorenzo",
    title = "{Two-loop QCD helicity amplitudes for $q\bar q \to W^\pm \gamma$ and $q\bar q \to Z^0 \gamma$}",
    eprint = "1112.1531",
    archivePrefix = "arXiv",
    primaryClass = "hep-ph",
    reportNumber = "ZU-TH-26-11, ZU-TH 26/11",
    doi = "10.1007/JHEP02(2012)004",
    journal = "JHEP",
    volume = "02",
    pages = "004",
    year = "2012"
}

@article{Anastasiou:2020sdt,
    author = "Anastasiou, Charalampos and Haindl, Rayan and Sterman, George and Yang, Zhou and Zeng, Mao",
    title = "{Locally finite two-loop amplitudes for off-shell multi-photon production in electron-positron annihilation}",
    eprint = "2008.12293",
    archivePrefix = "arXiv",
    primaryClass = "hep-ph",
    reportNumber = "YITP-SB-2020-25",
    doi = "10.1007/JHEP04(2021)222",
    journal = "JHEP",
    volume = "04",
    pages = "222",
    year = "2021"
}

@article{Anastasiou:2022eym,
    author = "Anastasiou, Charalampos and Sterman, George",
    title = "{Locally finite two-loop QCD amplitudes from IR universality for electroweak production}",
    eprint = "2212.12162",
    archivePrefix = "arXiv",
    primaryClass = "hep-ph",
    reportNumber = "YITP-SB-2022-43",
    doi = "10.1007/JHEP05(2023)242",
    journal = "JHEP",
    volume = "05",
    pages = "242",
    year = "2023"
}

@article{Anastasiou:2024xvk,
    author = "Anastasiou, Charalampos and Karlen, Julia and Sterman, George and Venkata, Aniruddha",
    title = "{Locally finite two-loop amplitudes for electroweak production through gluon fusion}",
    eprint = "2403.13712",
    archivePrefix = "arXiv",
    primaryClass = "hep-ph",
    doi = "10.1007/JHEP11(2024)043",
    journal = "JHEP",
    volume = "11",
    pages = "043",
    year = "2024"
}

@article{Anastasiou:2025cvy,
    author = "Anastasiou, Charalampos and Karlen, Julia and Sahoo, Roshni and Sterman, George and Vicini, Matilde",
    title = "{General finite two-loop amplitude integrand for photoproduction in quark annihilation}",
    eprint = "2509.07805",
    archivePrefix = "arXiv",
    primaryClass = "hep-ph",
    month = "9",
    year = "2025"
}

@article{Anastasiou:2026kpm,
    author = "Anastasiou, Charalampos and Karlen, Julia and Ma, Yao and Sterman, George",
    title = "{Local finiteness for real-virtual corrections to electroweak production in partonic collisions}",
    eprint = "2601.22936",
    archivePrefix = "arXiv",
    primaryClass = "hep-ph",
    month = "1",
    year = "2026"
}

@article{Bogoliubov:1957gp,
    author = "Bogoliubov, N. N. and Parasiuk, O. S.",
    title = "{On the Multiplication of the causal function in the quantum theory of fields}",
    doi = "10.1007/BF02392399",
    journal = "Acta Math.",
    volume = "97",
    pages = "227--266",
    year = "1957"
}

@article{Hepp:1966eg,
    author = "Hepp, Klaus",
    title = "{Proof of the Bogolyubov-Parasiuk theorem on renormalization}",
    doi = "10.1007/BF01773358",
    journal = "Commun. Math. Phys.",
    volume = "2",
    pages = "301--326",
    year = "1966"
}

@article{Zimmermann:1969jj,
    author = "Zimmermann, W.",
    title = "{Convergence of Bogolyubov's method of renormalization in momentum space}",
    doi = "10.1007/BF01645676",
    journal = "Commun. Math. Phys.",
    volume = "15",
    pages = "208--234",
    year = "1969"
}

@article{Catani:2008xa,
    author = "Catani, Stefano and Gleisberg, Tanju and Krauss, Frank and Rodrigo, German and Winter, Jan-Christopher",
    title = "{From loops to trees by-passing Feynman's theorem}",
    eprint = "0804.3170",
    archivePrefix = "arXiv",
    primaryClass = "hep-ph",
    reportNumber = "FERMILAB-PUB-08-092-T, IFIC-08-21, IPPP-08-22, SLAC-PUB-13218",
    doi = "10.1088/1126-6708/2008/09/065",
    journal = "JHEP",
    volume = "09",
    pages = "065",
    year = "2008"
}

@article{Aguilera-Verdugo:2019kbz,
    author = "Aguilera-Verdugo, J. Jes{\'u}s and Driencourt-Mangin, F{\'e}lix and Plenter, Judith and Ram{\'\i}rez-Uribe, Selomit and Rodrigo, Germ{\'a}n and Sborlini, Germ{\'a}n F. R. and Torres Bobadilla, William J. and Tracz, Szymon",
    title = "{Causality, unitarity thresholds, anomalous thresholds and infrared singularities from the loop-tree duality at higher orders}",
    eprint = "1904.08389",
    archivePrefix = "arXiv",
    primaryClass = "hep-ph",
    reportNumber = "IFIC/19-22",
    doi = "10.1007/JHEP12(2019)163",
    journal = "JHEP",
    volume = "12",
    pages = "163",
    year = "2019"
}

@article{Aguilera-Verdugo:2020set,
    author = "Aguilera-Verdugo, J. Jesus and Driencourt-Mangin, Felix and Hern{\'a}ndez-Pinto, Roger J. and Plenter, Judith and Ramirez-Uribe, Selomit and Renteria Olivo, Andres E. and Rodrigo, German and Sborlini, German F. R. and Torres Bobadilla, William J. and Tracz, Szymon",
    title = "{Open Loop Amplitudes and Causality to All Orders and Powers from the Loop-Tree Duality}",
    eprint = "2001.03564",
    archivePrefix = "arXiv",
    primaryClass = "hep-ph",
    reportNumber = "IFIC/20-02",
    doi = "10.1103/PhysRevLett.124.211602",
    journal = "Phys. Rev. Lett.",
    volume = "124",
    number = "21",
    pages = "211602",
    year = "2020"
}

@article{JesusAguilera-Verdugo:2020fsn,
    author = "Jes{\'u}s Aguilera-Verdugo, J. and Hern{\'a}ndez-Pinto, Roger J. and Rodrigo, Germ{\'a}n and Sborlini, German F. R. and Torres Bobadilla, William J.",
    title = "{Mathematical properties of nested residues and their application to multi-loop scattering amplitudes}",
    eprint = "2010.12971",
    archivePrefix = "arXiv",
    primaryClass = "hep-ph",
    reportNumber = "IFIC/20-30; DESY 20-172; MPP-2020-184",
    doi = "10.1007/JHEP02(2021)112",
    journal = "JHEP",
    volume = "02",
    pages = "112",
    year = "2021"
}

@article{Ramirez-Uribe:2022sja,
    author = "Ram{\'\i}rez-Uribe, Selomit and Hern{\'a}ndez-Pinto, Roger Jos{\'e} and Rodrigo, Germ{\'a}n and Sborlini, German F. R.",
    title = "{From Five-Loop Scattering Amplitudes to Open Trees with the Loop-Tree Duality}",
    eprint = "2211.03163",
    archivePrefix = "arXiv",
    primaryClass = "hep-ph",
    reportNumber = "IFIC/20-30",
    doi = "10.3390/sym14122571",
    journal = "Symmetry",
    volume = "14",
    number = "12",
    pages = "2571",
    year = "2022"
}

@article{Runkel:2019yrs,
    author = "Runkel, Robert and Sz{\H{o}}r, Zolt{\'a}n and Vesga, Juan Pablo and Weinzierl, Stefan",
    title = "{Causality and loop-tree duality at higher loops}",
    eprint = "1902.02135",
    archivePrefix = "arXiv",
    primaryClass = "hep-ph",
    doi = "10.1103/PhysRevLett.122.111603",
    journal = "Phys. Rev. Lett.",
    volume = "122",
    number = "11",
    pages = "111603",
    year = "2019",
    note = "[Erratum: Phys.Rev.Lett. 123, 059902 (2019)]"
}

@article{Capatti:2019ypt,
    author = "Capatti, Zeno and Hirschi, Valentin and Kermanschah, Dario and Ruijl, Ben",
    title = "{Loop-Tree Duality for Multiloop Numerical Integration}",
    eprint = "1906.06138",
    archivePrefix = "arXiv",
    primaryClass = "hep-ph",
    doi = "10.1103/PhysRevLett.123.151602",
    journal = "Phys. Rev. Lett.",
    volume = "123",
    number = "15",
    pages = "151602",
    year = "2019"
}

@article{Bodwin:1984hc,
    author = "Bodwin, Geoffrey T.",
    title = "{Factorization of the Drell-Yan Cross-Section in Perturbation Theory}",
    reportNumber = "ANL-HEP-PR-84-64-REV, ANL-HEP-PR-84-64",
    doi = "10.1103/PhysRevD.34.3932",
    journal = "Phys. Rev. D",
    volume = "31",
    pages = "2616",
    year = "1985",
    note = "[Erratum: Phys.Rev.D 34, 3932 (1986)]"
}

@article{Collins:1985ue,
    author = "Collins, John C. and Soper, Davison E. and Sterman, George F.",
    title = "{Factorization for Short Distance Hadron - Hadron Scattering}",
    reportNumber = "OITS-287",
    doi = "10.1016/0550-3213(85)90565-6",
    journal = "Nucl. Phys. B",
    volume = "261",
    pages = "104--142",
    year = "1985"
}

@book{Sterman:1993hfp,
    author = "Sterman, George F.",
    title = "{An Introduction to quantum field theory}",
    isbn = "978-0-521-31132-8",
    publisher = "Cambridge University Press",
    month = "8",
    year = "1993"
}

@inproceedings{Sterman:1995fz,
    author = "Sterman, George F.",
    title = "{Partons, factorization and resummation, TASI 95}",
    booktitle = "{Theoretical Advanced Study Institute in Elementary Particle Physics (TASI 95): QCD and Beyond}",
    eprint = "hep-ph/9606312",
    archivePrefix = "arXiv",
    reportNumber = "ITP-SB-96-4",
    pages = "327--408",
    month = "6",
    year = "1995"
}

@article{Sterman:2023xdj,
    author = "Sterman, George and Venkata, Aniruddha",
    title = "{Local infrared safety in time-ordered perturbation theory}",
    eprint = "2309.13023",
    archivePrefix = "arXiv",
    primaryClass = "hep-ph",
    reportNumber = "YITP-SB-2023-28",
    doi = "10.1007/JHEP02(2024)101",
    journal = "JHEP",
    volume = "02",
    pages = "101",
    year = "2024"
}

@article{Aguilera-Verdugo:2020kzc,
    author = "Aguilera-Verdugo, J. Jesus and Hernandez-Pinto, Roger J. and Rodrigo, German and Sborlini, German F. R. and Torres Bobadilla, William J.",
    title = "{Causal representation of multi-loop Feynman integrands within the loop-tree duality}",
    eprint = "2006.11217",
    archivePrefix = "arXiv",
    primaryClass = "hep-ph",
    reportNumber = "IFIC/20-27",
    doi = "10.1007/JHEP01(2021)069",
    journal = "JHEP",
    volume = "01",
    pages = "069",
    year = "2021"
}

@article{Ramirez-Uribe:2020hes,
    author = "Ram{\'\i}rez-Uribe, Selomit and Hern{\'a}ndez-Pinto, Roger J. and Rodrigo, German and Sborlini, Germ{\'a}n F. R. and Torres Bobadilla, William J.",
    title = "{Universal opening of four-loop scattering amplitudes to trees}",
    eprint = "2006.13818",
    archivePrefix = "arXiv",
    primaryClass = "hep-ph",
    reportNumber = "IFIC/20-29",
    doi = "10.1007/JHEP04(2021)129",
    journal = "JHEP",
    volume = "04",
    pages = "129",
    year = "2021"
}

@article{Sborlini:2021owe,
    author = "Sborlini, German F. R.",
    title = "{Geometrical approach to causality in multiloop amplitudes}",
    eprint = "2102.05062",
    archivePrefix = "arXiv",
    primaryClass = "hep-ph",
    reportNumber = "DESY-21-017, DESY 21-017",
    doi = "10.1103/PhysRevD.104.036014",
    journal = "Phys. Rev. D",
    volume = "104",
    number = "3",
    pages = "036014",
    year = "2021"
}

@article{TorresBobadilla:2021ivx,
    author = "Torres Bobadilla, William J.",
    title = "{Loop-tree duality from vertices and edges}",
    eprint = "2102.05048",
    archivePrefix = "arXiv",
    primaryClass = "hep-ph",
    reportNumber = "MPP-2021-14",
    doi = "10.1007/JHEP04(2021)183",
    journal = "JHEP",
    volume = "04",
    pages = "183",
    year = "2021"
}

@article{Bobadilla:2021pvr,
    author = "Bobadilla, William J. Torres",
    title = "{Lotty {\textendash} The loop-tree duality automation}",
    eprint = "2103.09237",
    archivePrefix = "arXiv",
    primaryClass = "hep-ph",
    reportNumber = "MPP-2021-11",
    doi = "10.1140/epjc/s10052-021-09235-0",
    journal = "Eur. Phys. J. C",
    volume = "81",
    number = "6",
    pages = "514",
    year = "2021"
}

@article{Benincasa:2021qcb,
    author = "Benincasa, Paolo and Bobadilla, William J. Torres",
    title = "{Physical representations for scattering amplitudes and the wavefunction of the universe}",
    eprint = "2112.09028",
    archivePrefix = "arXiv",
    primaryClass = "hep-th",
    reportNumber = "MPP-2021-186",
    doi = "10.21468/SciPostPhys.12.6.192",
    journal = "SciPost Phys.",
    volume = "12",
    number = "6",
    pages = "192",
    year = "2022"
}

@article{Kromin:2022txz,
    author = "Kromin, Sascha and Schwanemann, Niklas and Weinzierl, Stefan",
    title = "{Amplitudes within causal loop-tree duality}",
    eprint = "2208.01060",
    archivePrefix = "arXiv",
    primaryClass = "hep-th",
    reportNumber = "MITP/22-071",
    doi = "10.1103/PhysRevD.106.076006",
    journal = "Phys. Rev. D",
    volume = "106",
    number = "7",
    pages = "076006",
    year = "2022"
}

@article{Capatti:2020ytd,
    author = "Capatti, Zeno and Hirschi, Valentin and Kermanschah, Dario and Pelloni, Andrea and Ruijl, Ben",
    title = "{Manifestly Causal Loop-Tree Duality}",
    eprint = "2009.05509",
    archivePrefix = "arXiv",
    primaryClass = "hep-ph",
    month = "9",
    year = "2020"
}

@article{Capatti:2022mly,
    author = "Capatti, Zeno",
    title = "{Exposing the threshold structure of loop integrals}",
    eprint = "2211.09653",
    archivePrefix = "arXiv",
    primaryClass = "hep-th",
    doi = "10.1103/PhysRevD.107.L051902",
    journal = "Phys. Rev. D",
    volume = "107",
    number = "5",
    pages = "L051902",
    year = "2023"
}

@article{Capatti:2023shz,
    author = "Capatti, Zeno",
    title = "{Derivation of the Cross-Free Family representation for the box diagram}",
    eprint = "2311.14374",
    archivePrefix = "arXiv",
    primaryClass = "hep-ph",
    doi = "10.22323/1.432.0027",
    journal = "PoS",
    volume = "RADCOR2023",
    pages = "027",
    year = "2024"
}

@article{Kermanschah:2021wbk,
    author = "Kermanschah, Dario",
    title = "{Numerical integration of loop integrals through local cancellation of threshold singularities}",
    eprint = "2110.06869",
    archivePrefix = "arXiv",
    primaryClass = "hep-ph",
    doi = "10.1007/JHEP01(2022)151",
    journal = "JHEP",
    volume = "01",
    pages = "151",
    year = "2022"
}

@article{Vicini:2024ecf,
    author = "Vicini, Matilde and Kermanschah, Dario",
    title = "{Numerical integration of the double- and triple-box integrals using threshold subtraction}",
    eprint = "2407.21511",
    archivePrefix = "arXiv",
    primaryClass = "hep-ph",
    doi = "10.22323/1.467.0078",
    journal = "PoS",
    volume = "LL2024",
    pages = "078",
    year = "2024"
}

@article{Soper:1999xk,
    author = "Soper, Davison E.",
    title = "{Techniques for QCD calculations by numerical integration}",
    eprint = "hep-ph/9910292",
    archivePrefix = "arXiv",
    doi = "10.1103/PhysRevD.62.014009",
    journal = "Phys. Rev. D",
    volume = "62",
    pages = "014009",
    year = "2000"
}

@article{Becker:2011vg,
    author = "Becker, Sebastian and Goetz, Daniel and Reuschle, Christian and Schwan, Christopher and Weinzierl, Stefan",
    title = "{NLO results for five, six and seven jets in electron-positron annihilation}",
    eprint = "1111.1733",
    archivePrefix = "arXiv",
    primaryClass = "hep-ph",
    doi = "10.1103/PhysRevLett.108.032005",
    journal = "Phys. Rev. Lett.",
    volume = "108",
    pages = "032005",
    year = "2012"
}

@article{Becker:2012aqa,
    author = "Becker, Sebastian and Reuschle, Christian and Weinzierl, Stefan",
    title = "{Efficiency Improvements for the Numerical Computation of NLO Corrections}",
    eprint = "1205.2096",
    archivePrefix = "arXiv",
    primaryClass = "hep-ph",
    doi = "10.1007/JHEP07(2012)090",
    journal = "JHEP",
    volume = "07",
    pages = "090",
    year = "2012"
}

@article{Buchta:2015wna,
    author = "Buchta, Sebastian and Chachamis, Grigorios and Draggiotis, Petros and Rodrigo, German",
    title = "{Numerical implementation of the loop{\textendash}tree duality method}",
    eprint = "1510.00187",
    archivePrefix = "arXiv",
    primaryClass = "hep-ph",
    reportNumber = "IFIC-15-69",
    doi = "10.1140/epjc/s10052-017-4833-6",
    journal = "Eur. Phys. J. C",
    volume = "77",
    number = "5",
    pages = "274",
    year = "2017"
}

@article{Capatti:2019edf,
    author = "Capatti, Zeno and Hirschi, Valentin and Kermanschah, Dario and Pelloni, Andrea and Ruijl, Ben",
    title = "{Numerical Loop-Tree Duality: contour deformation and subtraction}",
    eprint = "1912.09291",
    archivePrefix = "arXiv",
    primaryClass = "hep-ph",
    doi = "10.1007/JHEP04(2020)096",
    journal = "JHEP",
    volume = "04",
    pages = "096",
    year = "2020"
}

@article{Lepage:2020tgj,
    author = "Lepage, G. Peter",
    title = "{Adaptive multidimensional integration: VEGAS enhanced}",
    eprint = "2009.05112",
    archivePrefix = "arXiv",
    primaryClass = "physics.comp-ph",
    doi = "10.1016/j.jcp.2021.110386",
    journal = "J. Comput. Phys.",
    volume = "439",
    pages = "110386",
    year = "2021"
}

@article{Lepage:1977sw,
    author = "Lepage, G. Peter",
    title = "{A New Algorithm for Adaptive Multidimensional Integration}",
    reportNumber = "SLAC-PUB-1839-REV, SLAC-PUB-1839",
    doi = "10.1016/0021-9991(78)90004-9",
    journal = "J. Comput. Phys.",
    volume = "27",
    pages = "192",
    year = "1978"
}

@article{Hahn:2004fe,
    author = "Hahn, T.",
    title = "{CUBA: A Library for multidimensional numerical integration}",
    eprint = "hep-ph/0404043",
    archivePrefix = "arXiv",
    reportNumber = "MPP-2004-40",
    doi = "10.1016/j.cpc.2005.01.010",
    journal = "Comput. Phys. Commun.",
    volume = "168",
    pages = "78--95",
    year = "2005"
}

@article{Kuipers:2013pba,
    author = "Kuipers, J. and Ueda, T. and Vermaseren, J. A. M.",
    title = "{Code Optimization in FORM}",
    eprint = "1310.7007",
    archivePrefix = "arXiv",
    primaryClass = "cs.SC",
    reportNumber = "NIKHEF-2013-036, TTP13-031, SFB-CPP-13-80",
    doi = "10.1016/j.cpc.2014.08.008",
    journal = "Comput. Phys. Commun.",
    volume = "189",
    pages = "1--19",
    year = "2015"
}

@article{Ruijl:2013epa,
    author = "Ruijl, Ben and Vermaseren, Jos and Plaat, Aske and Herik, Jaap van den",
    title = "{Combining Simulated Annealing and Monte Carlo Tree Search for Expression Simplification}",
    eprint = "1312.0841",
    archivePrefix = "arXiv",
    primaryClass = "cs.AI",
    month = "12",
    year = "2013"
}

@article{Davies:2026cci,
    author = "Davies, J. and Kaneko, T. and Marinissen, C. and Ueda, T. and Vermaseren, J. A. M.",
    title = "{FORM Version 5.0}",
    eprint = "2601.19982",
    archivePrefix = "arXiv",
    primaryClass = "hep-ph",
    month = "1",
    year = "2026"
}

@article{Kilian:2009wy,
    author = "Kilian, Wolfgang and Kleinschmidt, Tobias",
    title = "{Numerical Evaluation of Feynman Loop Integrals by Reduction to Tree Graphs}",
    eprint = "0912.3495",
    archivePrefix = "arXiv",
    primaryClass = "hep-ph",
    reportNumber = "IPPP-09-97, DCTP-09-194, EDINBURGH-2009-17, SI-HEP-2009-16",
    month = "12",
    year = "2009"
}

@article{Becchetti:2025rrz,
    author = "Becchetti, Matteo and Coro, Federico and Nega, Christoph and Tancredi, Lorenzo and Wagner, Fabian J.",
    title = "{Analytic two-loop amplitudes for $ q\overline{q}\to \gamma \gamma $ and gg {\textrightarrow} {\ensuremath{\gamma}}{\ensuremath{\gamma}} mediated by a heavy-quark loop}",
    eprint = "2502.00118",
    archivePrefix = "arXiv",
    primaryClass = "hep-ph",
    reportNumber = "TUM-HEP 1554/25",
    doi = "10.1007/JHEP06(2025)033",
    journal = "JHEP",
    volume = "06",
    pages = "033",
    year = "2025"
}

@article{Ahmed:2025osb,
    author = "Ahmed, Taushif and Chakraborty, Amlan and Chaubey, Ekta and Kaur, Mandeep",
    title = "{Two-loop helicity amplitudes for diphoton production with massive quark loop}",
    eprint = "2502.03282",
    archivePrefix = "arXiv",
    primaryClass = "hep-ph",
    reportNumber = "TIF-UNIMI-2025-4, BONN-TH-2025-01, DESY-25-017",
    doi = "10.1007/JHEP12(2025)106",
    journal = "JHEP",
    volume = "12",
    pages = "106",
    year = "2025"
}

@article{Lai:2010vv,
    author = "Lai, Hung-Liang and Guzzi, Marco and Huston, Joey and Li, Zhao and Nadolsky, Pavel M. and Pumplin, Jon and Yuan, C. -P.",
    title = "{New parton distributions for collider physics}",
    eprint = "1007.2241",
    archivePrefix = "arXiv",
    primaryClass = "hep-ph",
    reportNumber = "MSUHEP-100707, SMU-HEP-10-10",
    doi = "10.1103/PhysRevD.82.074024",
    journal = "Phys. Rev. D",
    volume = "82",
    pages = "074024",
    year = "2010"
}

@article{Buckley:2014ana,
    author = {Buckley, Andy and Ferrando, James and Lloyd, Stephen and Nordstr{\"o}m, Karl and Page, Ben and R{\"u}fenacht, Martin and Sch{\"o}nherr, Marek and Watt, Graeme},
    title = "{LHAPDF6: parton density access in the LHC precision era}",
    eprint = "1412.7420",
    archivePrefix = "arXiv",
    primaryClass = "hep-ph",
    reportNumber = "GLAS-PPE-2014-05, MCNET-14-29, IPPP-14-111, DCPT-14-222",
    doi = "10.1140/epjc/s10052-015-3318-8",
    journal = "Eur. Phys. J. C",
    volume = "75",
    pages = "132",
    year = "2015"
}

@article{Grazzini:2017mhc,
    author = "Grazzini, Massimiliano and Kallweit, Stefan and Wiesemann, Marius",
    title = "{Fully differential NNLO computations with MATRIX}",
    eprint = "1711.06631",
    archivePrefix = "arXiv",
    primaryClass = "hep-ph",
    reportNumber = "ZU-TH-30-17, CERN-TH-2017-232, ZU-TH 30/17",
    doi = "10.1140/epjc/s10052-018-5771-7",
    journal = "Eur. Phys. J. C",
    volume = "78",
    number = "7",
    pages = "537",
    year = "2018"
}

@article{Grazzini:2013bna,
    author = "Grazzini, Massimiliano and Kallweit, Stefan and Rathlev, Dirk and Torre, Alessandro",
    title = "{$Z\gamma$ production at hadron colliders in NNLO QCD}",
    eprint = "1309.7000",
    archivePrefix = "arXiv",
    primaryClass = "hep-ph",
    reportNumber = "ZU-TH-21-13",
    doi = "10.1016/j.physletb.2014.02.037",
    journal = "Phys. Lett. B",
    volume = "731",
    pages = "204--207",
    year = "2014"
}

@article{Grazzini:2015nwa,
    author = "Grazzini, Massimiliano and Kallweit, Stefan and Rathlev, Dirk",
    title = "{$W\gamma$ and $Z\gamma$ production at the LHC in NNLO QCD}",
    eprint = "1504.01330",
    archivePrefix = "arXiv",
    primaryClass = "hep-ph",
    reportNumber = "ZU-TH-04-15, MITP-15-021",
    doi = "10.1007/JHEP07(2015)085",
    journal = "JHEP",
    volume = "07",
    pages = "085",
    year = "2015"
}

@article{Campbell:2017aul,
    author = "Campbell, John M. and Neumann, Tobias and Williams, Ciaran",
    title = "{$Z\gamma$ Production at NNLO Including Anomalous Couplings}",
    eprint = "1708.02925",
    archivePrefix = "arXiv",
    primaryClass = "hep-ph",
    reportNumber = "FERMILAB-PUB-17-303-T",
    doi = "10.1007/JHEP11(2017)150",
    journal = "JHEP",
    volume = "11",
    pages = "150",
    year = "2017"
}

@misc{havana,
  author       = {Ruijl, Ben},
  title        = {havana: A Rust Monte Carlo integrator that supports discrete and continuous dimensions and nested grids},
  year         = {2025},
  howpublished = {\url{https://github.com/benruijl/havana}},
  note         = {GitHub repository},
}

@misc{spenso,
  author       = {Fraaije, Mathijs and Hirschi, Valentin and Huber, Lucien and Ruijl,Ben},
  title        = {Spenso},
  year         = {2024},
  howpublished = {\url{https://github.com/alphal00p/spenso/releases/tag/spynso3-v0.1.0}},
  note         = {GitHub tag v0.1.0},
}

@software{symbolica,
  author       = {Ruijl, Ben},
  title        = {Symbolica},
  month        = sep,
  year         = 2025,
  publisher    = {Ruijl Research},
  version      = {0.18.0},
  doi          = {10.5281/zenodo.17054381},
  url          = {https://doi.org/10.5281/zenodo.17054381},
  swhid        = {swh:1:dir:7d96543a9deb63b49aa5896fa88fb84e8c6968be
                   ;origin=https://doi.org/10.5281/zenodo.15040847;vi
                   sit=swh:1:snp:6c95ac4bb713c8b25e38d14ed172fde43f68
                   9769;anchor=swh:1:rel:c9e3f50850a5330f4fc5dda10a38
                   f914fd2dd0a0;path=/
                  },
}

@misc{twofloat_dev,
  author       = {Tribick, Andrew and Cruz, José Luis  and Pelloni, Andrea},
  title        = {twofloat},
  year         = {2026},
  howpublished = {\url{https://github.com/apelloni/twofloat/tree/dev}},
  note         = {GitHub fork of the original repository, dev branch},
}

@article{double_double,
author = {Joldes, Mioara and Muller, Jean-Michel and Popescu, Valentina},
title = {Tight and Rigorous Error Bounds for Basic Building Blocks of Double-Word Arithmetic},
year = {2017},
issue_date = {June 2018},
publisher = {Association for Computing Machinery},
address = {New York, NY, USA},
volume = {44},
number = {2},
issn = {0098-3500},
url = {https://doi.org/10.1145/3121432},
doi = {10.1145/3121432},
journal = {ACM Trans. Math. Softw.},
month = oct,
articleno = {15res},
numpages = {27}
}
